# Observation of surface plasmon polaritons in 2D electron gas of surface electron accumulation in InN nanostructures

*Kishore K. Madapu,*[*,†] *A. K. Sivadasan,*[†,#] *Madhusmita Baral,*[‡]  *Sandip Dhara*[*,†]

[†]Nanomaterials Characterization and Sensors Section, Surface and Nanoscience Division, Indira Gandhi Centre for Atomic Research, Homi Bhabha National Institute, Kalpakkam-603102, India

[‡]Synchrotron Utilization Section, Raja Ramanna Centre for Advanced technology, Homi Bhabha National Institute, Indore-452013, India

Real space imaging of propagating surface plasmon polaritons having the wavelength in the 500 nm, for the first time, indicating InN as a promising alternate for the 2D low loss plasmonic material in the THz region.

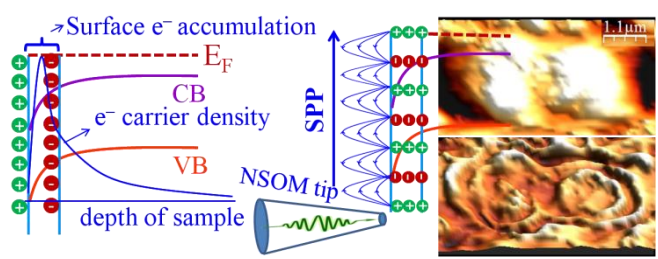

## Abstract

Recently, heavily doped semiconductors are emerging as an alternate for low loss plasmonic materials. InN, belonging to the group III nitrides, possesses the unique property of surface electron accumulation (SEA) which provides two dimensional electron gas (2DEG) system. In this report, we demonstrated the surface plasmon properties of InN nanoparticles originating from SEA using the real space mapping of the surface plasmon fields for the first time. The SEA is confirmed by Raman studies which are further corroborated by photoluminescence and photoemission spectroscopic studies. The frequency of 2DEG corresponding to SEA is found to be in the THz region.  The periodic fringes are observed in the near–field scanning optical microscopic images of InN nanostructures. The observed fringes are attributed to the interference of propagated and back reflected surface plasmon polaritons (SPPs).  The observation of SPPs is solely attributed to the 2DEG corresponding to the SEA of InN. In addition, resonance kind of behavior with the enhancement of the near-field intensity is observed in the near-field images of InN nanostructures. Observation of SPPs indicates that InN with SEA can be a promising THz plasmonic material for the light confinement.

# 1. Introduction

In the last decade surface plasmons, categorized as plasmonics, is widely studied because of its potentiality in the miniaturization of photonic devices and other widespread applications in the multiple fields like biological and chemical sensing [1–5]. Surface plasmon is collective oscillations of conduction band electrons excited by the electromagnetic waves. The surface plasmons have the two fundamental excitations such as localized surface plasmon resonance (LSPR) and propagating surface plasmon polaritons (SPPs). The SPPs are propagating electromagnetic waves along the metal and dielectric interface, which are confined along the interface region. The most spectacular application of the SPP is its ability to confine and guide the light in the sub-wavelength structures by overcoming the Abbe's diffraction limit [6, 7]. So far, the most studied plasmonic materials are Au and Ag, which show LSPR phenomenon in the infra-red (IR) to the visible region. Even though major studied plasmonic materials are noble metals, however, use of metal nanoparticles is having its own limitations because of the plasmonic losses. In this context, low-loss plasmonic metamaterials [8], and semiconductors with sufficiently high carrier density can be the alternative for metallic nanostructures in the field of plasmonics [9, 10]. In case of semiconductors, the LSPR frequency can be achieved up to near infrared range [11]. Moreover, plasmonics based on metals is static in nature. The carrier concentration, and subsequently the plasmonic properties of semiconductors can be readily altered with doping in semiconductors. Thus, the semiconductor nanoparticles with varying carrier concentration can offer the dynamic plasmonic nature [12]. Till now the metal chalcogenides and metal oxides are studied for their plasmonic properties [13, 14]. Recently, two-dimensional (2D) plasmons in graphene have generated lot of curiosity because of its terahertz (THz) resonance frequency [15–19]. In case of 3D plasmonic material, the plasmon frequency ($\omega_p$) is scaled with square root of carrier density of the system. On the other hand, $\omega_p$ depends on the in-plane wave vector as well as areal carrier density in case of the 2D plasmonics [20–22]. The plasmon frequency of two-dimensional electron gases (2DEG) is written as

$$\omega_p^2 = \frac{n_s e^2}{m^*} \frac{q}{\varepsilon_1 + \varepsilon_2 \coth(qd)} \qquad (1)$$

where $n_s$ is the areal carrier density of 2D electron gas, $m^*$ effective mass of electron, $\varepsilon_1$ and $\varepsilon_2$ are the dielectric constants of the substrate and surrounding media, $q$ is the in-plane wave vector, and $d$ is the thickness of the two-dimensional electron system.

The plasmonics of the graphene [15-19] is attributed the 2D mass-less Dirac Fermions. The plasmon resonance frequency of graphene is observed in the terahertz (THz) region [15]. The carrier density in the graphene, and subsequently the plasmon resonance frequency is readily varied using applied gate voltage. F. Zhe *et al.* [17], reported 2D plasmon excitations of Dirac Fermions using the infrared (IR) nanoscopy. Later, real space imaging of plasmon fields of graphene is carried out using the scattering-type scanning near-field optical microscopy (s-SNOM) [18, 19]. Periodic fringes are observed in the s-SNOM images, which are attributed to the SPPs of Dirac Fermions. Even though graphene plasmon frequency is in the range of THz, the wavelength of SPPs (~ 260 nm) is observed forty times less than the excitation wavelength. The extreme compression of the plasmon wavelength is attributed to the two-dimensionality and unique conductance of the graphene. In addition, shortening of the SPP wavelength is governed by the fine-structure constant ($\alpha = 1/137$). The relationship between excitation and plasmon wavelength is given by [19],

$$\lambda_p/\lambda_o \approx [4\alpha/((\varepsilon+1))] (E_F/E_p) \qquad (2)$$

where $\lambda_p$, $\lambda_o$ are the plasmon and excited wavelength respectively; $E_F$ and $E_p$ are Fermi energy and plasmon energy, respectively, and $\varepsilon$ is the dielectric constant of the substrate.

2D plasmons can also be observed in the semiconductor inversion layers such as Si and GaAs [20–22]. Recently, Dirac plasmons are also observed in the topological insulators because of the downward bending near the surface region [23]. However, materials like InAs, InN possess the inherent surface electron accumulation (SEA) which can act as a 2DEG. Among these, InN is reported to possess the higher amount of sheet carrier density as compared to that of InAs [24]. In this context, InN is the material which belongs to the III-nitride family and is having the unique property of possessing SEA [25–28]. InN attained a lot of interest in the research community because of the SEA, and superior electronic properties. The lowest effective electron mass and highest saturation velocity, as compared to other III-nitride semiconductors are the reasons for superior electronic properties of InN [29]. As a result, low field mobility of InN can reach up to 10,000 $cm^2V^{-1}s^{-1}$ [30]. The origin of the SEA is attributed to the location of the branch point energy ($E_{BP}$), which is the transition energy level of the donor to acceptor type states. In case of InN, the $E_{BP}$ is located above the conduction band minima; consequently there can be donor-type surface states in the conduction band. These donor-type surface states discharge their electrons into the conduction band by acquiring a positive charge. In order to compensate the positive charge of the surface states, electrons are accumulated near surface region. Moreover, downward band bending takes place near the surface region because of the positive charge of surface states [25–28]. The electrons in the SEA behaves like 2DEG which is confined in the direction perpendicular to the surface and free to move along the surface [26]. The sheet carrier density of the SEA is readily tuned using the growth parameters, [26] as well as external doping [27]. T. V. Subina *et al.* [32, 33], have reported the plasmonic properties of nanocomposites comprising of metallic In nanoparticles embedded in the InN matrix (In-InN). It was observed that optical properties of pure InN and In-InN composites are significantly different. The optical properties such as scattering, absorption, and emission of nanocomposites are controlled by Mie resonances of In nanoparticles. Mie resonance of In nanoparticles in InN matrix is observed in the energy range of 0.7-1 eV. Recently, THz emission from the InN films was also observed [34, 35]. The THz emission is attributed to the emission from the SPPs with the random grating coupling [34, 35]. Here, the SPPs are generated because of the 2DEG in the SEA.

In the present report, we explore the surface plasmon properties of the InN nanostructures using the real space mappingof the surface plasmon fields, for the first time, using near-field scanning optical microscopy (NSOM) technique. In this study, we have selected the high optical quality InN nanoparticles grown at different temperatures in the atmospheric chemical vapor deposition (APCVD) technique. The SEA of InN nanostructures is confirmed by Raman spectroscopy, which is further corroborated by photoluminescence (PL) and photoemission spectroscopic (PES) analyses. Tuning of the surface sheet carrier density of SEA is achieved by varying the growth temperature. The NSOM images show the presence of the periodic fringes because of the generation of SPPs originating from 2DEG corresponding to SEA.

## 2. Experimental details

### 2.1 Sample preparation

InN nanostructures were grown on crystalline (0001) oriented $Al_2O_3$ substrate at three different temperatures of 580 $^o$C (sample A), 620 $^o$C (sample B) and 630 $^o$C (sample C) using the APCVD technique. The experimental details were detailed elsewhere [36].

### 2.2 Characterisation techniques

Morphology of nanostructures was studied by field emission scanning electron microscope (FESEM; AURIGA, Zeiss). Vibrational properties were studied using the Raman spectroscopy (inVia, Renishaw, UK) with the excitation of 514.5 nm laser in the backscattering geometry. Subsequently, the scattered light was detected using a charge coupled device (CCD) and a 1800 gr·mm$^{-1}$ grating for monochromatization. The PL measurements were also carried out with same setup using the 785 nm laser excitation. Emitted luminescence was monochromatized using a 600 gr·mm$^{-1}$ grating and was collected using a single channel InGaAs photodetector in the backscattering geometry. Structural studies using glancing incidence X-ray diffraction (GIXRD; Bruker, D8 Discover) was performed using Cu $K_\alpha$ ($\lambda$= 1.5406 Å) line (supporting information (SI), Figure S1). Photoemission measurements were carried out at Raja Ramanna Centre for Advanced Technology (RRCAT), Indore, India. Valence band (VB) spectra were recorded using a PHOIBOS 150 hemispherical analyzer with monochromatic He-I source (21.218 eV). Vacuum inside the analysis chamber during experiment was ~8×10$^{-11}$ mbar. The samples A, B were sputtered *in situ* by Ar$^+$ with beam energy of 500 eV and current of 5 mA for 2 minutes. Sample C was sputtered with a beam energy of 500 eV and current of 3 mA for 20 sec.

The surface plasmons properties of InN nanostructures were explored using near-field imaging with the aid of NSOM technique (MultiView 4000, Nanonics Imaging Ltd., Israel). In the present study, the excitation of SPPs and their intensity distribution were imaged using the apertured NSOM probe in the reflection mode (schematic in SI, Fig. S2). In this particular study, 532 nm laser excitation and a ~100 nm Cr-Au metal coated apertured tip was used in the near-field. The MultiView 4000 uses normal force tuning fork technology with a high Q factor and phase feedback to allow control of the probe/sample separation. Scattered light was collected with confocal detection using 50 X objective lens of the numerical aperture ~0.42. The scattered light was detected using an avalanche photodiode (APD) single photon counter (Sens-Tech, Model DM0087C). The AFM scanning was carried in intermittent (tapping) mode with the raster scanning using the fast and slow axis movements. This tracking mode enabled us to get the trace and re-trace images. In the present report, all scans were carried out with a scan rate of 20 ms/point and the integration time of photon counter was 20 ms. Thus, the simultaneous topographic image was also acquired along with the intensity distribution of surface plasmon scattering. As a result of slow scan rates of 20 ms/point, artifacts can be observed in AFM and NSOM images because of the sample drift. However, scan rate is limited by the integration time of the photon counter. The scan rate should be greater than or equal to the photon counter integration time for a meaningful measurement.

## 3. Results and discussion

### 3.1 Optical and correlated surface electronic properties

InN nanostructures, grown on the sapphire substrates using the APCVD technique with metallic In and reactive $NH_3$ as precursors at three different temperatures, 580 $^o$C (sample A), 620 $^o$C (sample B) and 630 $^o$C (sample C) are used for the study. Figures 1(a)−1(c) show the typical morphology of nanostructures corresponding to samples A, B, and C, respectively. The FESEM microscopic study shows that nanostructures are devoid of any particular shape and are having complete random morphology.

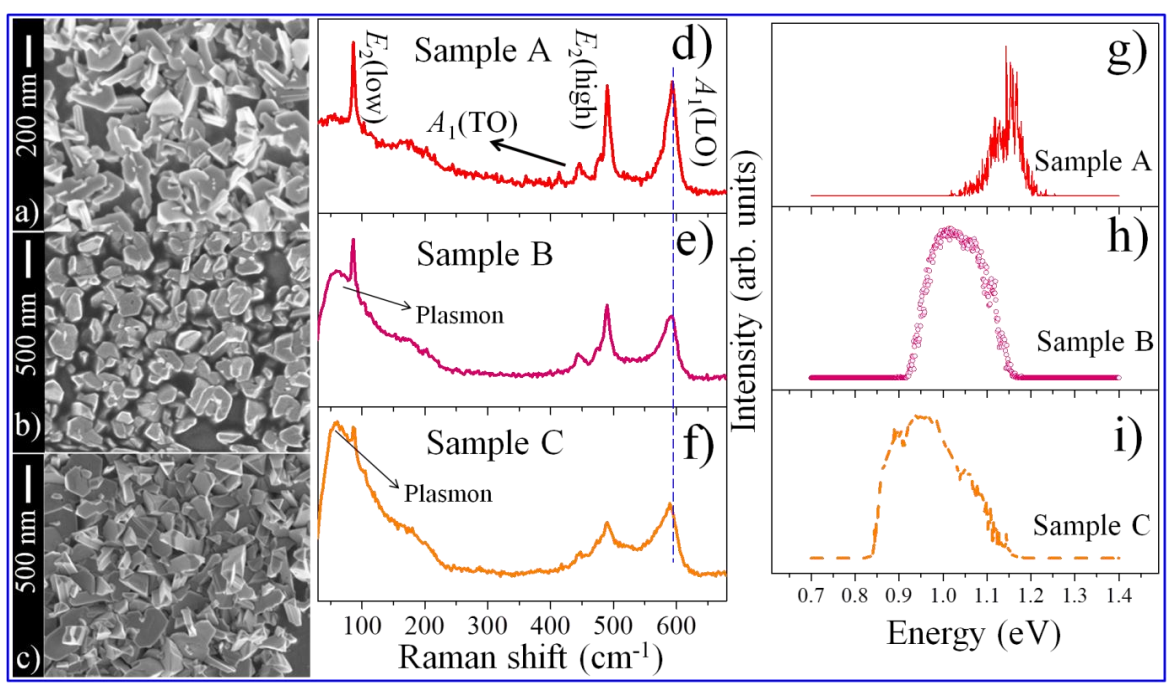

**Figure 1.** (a)-(c) FESEM micrographs showing the random nanostructures grown at 580 $^o$C (sample A), 620 $^o$C (sample B) and 630 $^o$C (sample C); (d)-(f) their corresponding Raman spectra, and (g)-(i) PL spectra. Vertical dashed line in (d)-(f) indicating the red shift in the $A_1$(LO) phonon mode with increasing growth temperature.

As discussed earlier, InN possesses SEA which can influence its vibrational and optical properties. The longitudinal optical (LO) phonon modes create the microscopic electric field during the vibration, and as a result, the LO phonon modes are expected to couple with free carrier density through the Fröhlich interaction. The coupling of LO phonon mode and carrier density leads to the observation of plasmon-phonon coupling modes (PCPs) in the vibrational spectrum [36, 37]. Room temperature Raman spectra of samples A, B and C are shown in Figures 1(d)−1(f), respectively. The distinct peaks are observed around the wave numbers of 87, 445, 490 and 580 cm$^{-1}$ which correspond to the symmetry allowed modes of $E_2$(low), transverse optic $A_1$(TO), $E_2$(high) and $A_1$(LO) phonon of the wurtzite InN, respectively. The redshift in the $A_1$(LO) phonon mode is observed with increasing the growth temperature i.e. sample A to sample C. Interestingly, instead of coupling modes, a difference in the line shape of the $A_1$(LO) phonon mode with asymmetric broadening is observed in samples B and C as compared to that for the sample A. The asymmetric line shape of the $A_1$(LO) mode is attributed to the Fano interference between carrier density and the $A_1$(LO) phonon modes. Usually, Fano line shape arises because of the interference between discrete phonon modes and background continuum electron transitions in the system [38]. The asymmetric line shape of $A_1$(LO) phonon mode elucidates the fact that samples B and C may be having higher carrier density in the system as compared to that for the sample A. It may be understood from the following facts that the probability of formation of N vacancies increases with increasing growth temperature because of the low thermal stability of InN [36]. At the same time, N vacancies contribute to the background carrier density [29], leading to asymmetric broadening of $A_1$(LO) phonon mode. For the sake of clarity, similar kind of asymmetric line shape of $A_1$(LO) phonon mode, as well as increase in FWHM, are observed for all samples grown above the 620 °C (not shown in picture).

In addition to the asymmetric broadening of $A_1$(LO) phonon, there is a striking difference in the Raman spectra (Figures 1(d)−1(f)) near the low wavenumber region. In case of samples B and C, there is a tiny peak emerging around 57 cm$^{-1}$. As the growth temperature increases this peak is strengthened (sample C), which is clearly visible in the spectra. The low-frequency spectral feature may be attributed to the plasmon excitation in the InN nanostructures system. However, the bulk plasmon frequency of the present nanostructures with the carrier density of ~$2.53 \times 10^{19}$ cm$^{-3}$ is calculated to be 2168 cm$^{-1}$. The carrier density is estimated from the Burstein-Moss (BM) shift in PL spectra (discussed in the subsequent section). In addition, the LSPR frequency of the In clusters in the InN matrix, if at all present, is reported to be in the range of 0.5−1 eV (4033−8100 cm$^{-1}$) [32, 33]. Thus, the observed low-frequency spectral feature can neither be attributed to the bulk plasmon frequency nor as the LSPR of In clusters of InN matrix. The low-frequency peak may originate because of 2D plasmonic oscillations of SEA. It was earlier reported that 2D plasmonic oscillations can be probed using the Raman spectroscopy [21, 22]. The condition for the excitation of the 2D plasmons in layered structure is $k_\| d \ll 1$, where $k_\|$ is the in-plane wave vector and *d* is the thickness of the layered structure [21]. In backscattering geometry, the in-plane wave vectors can be written as the $k_\| = (2\pi/\lambda)(\cos\theta - \sin\theta)$, where θ is the angle between the incident beam and normal to the surface [21]. Here, the angle between the incident and scattered light is assumed as 90 degrees. Because of the random alignment nano-crystals with respect to incident laser, the minimum and the maximum in-plane wave vector transfer occurs at angles of 45 and 0 degrees, respectively. The width of the SEA of the InN is in the range 4-10 nm [36], and the maximum in-plane wave vector is $\approx 2\pi/\lambda$ (θ≈0) which fulfil the condition for the 2D plasmon excitation. As a result, observed low-frequency mode is assigned as 2D plasmon peak. To corroborate the assignment of this peak, we calculated the 2D plasmon frequency of SEA for InN using the Equation (1). Figure 2(a) shows the variation of the plasma frequency of with respect to the areal carrier density. For the calculation of plasmon frequency, the bulk InN is taken as the substrate ($\varepsilon_1 = 14$), width of the SEA is 6 nm and *m\** as $0.13 m_o$ ($m_o$, rest mass of electron) in SEA region [35].

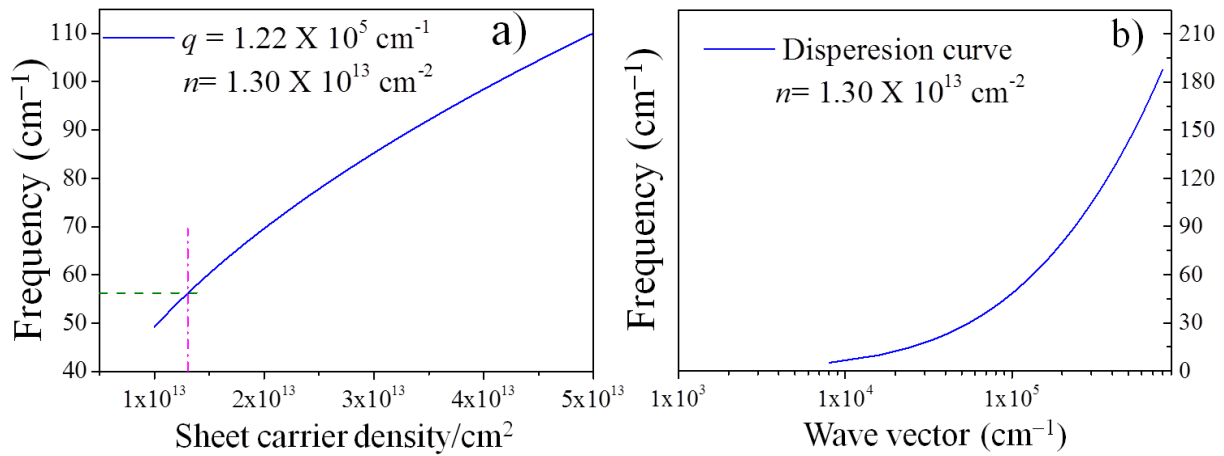

**Figure 2.** a) Sheet carrier density dependent 2D plasmon frequency of 2DEG corresponding to SEA of InN and its b) Dispersion curve.

The calculated plasma frequency (Figure 2a) is of the order of the observed low-frequency peak in the Raman spectra. The observed plasma frequency (57 cm$^{-1}$ ~ 1.7 THz) in the Raman spectra (Figures 1(b), 1(c)) corresponds to the areal carrier density $1.3 \times 10^{13}$ cm$^{-2}$, which closely matches with earlier reported values [39]. The bulk carrier density calculated from the BM shift, which is discussed in subsequent section, is also supportive to our assignment of the 2D plasmon peak. In addition, Figure 2(b) shows the dispersion curve of 2D plasmons for InN for a given carrier density $1.3 \times 10^{13}$ cm$^{-2}$. Moreover, it is observed that the plasmon frequency blue shifts as the growth temperature increases above 620 °C (SI,

Figure S3). As discussed earlier, carrier density is increased as there is an increase in growth temperature. As a result, 2D plasmon frequency increases with increasing growth temperature. The plasmonic property is corroborated by the PL emission studies of these nanostructures. The integrated PL intensity of high temperature grown samples (B and C) is higher than that of the low temperature grown sample A (Figures 1(g)-1(i)), indicating possibility of higher amount of SEA for InN nanostructures in the formers [40]. The broadening of band edge emission is attributed to the combination of radiative recombination processes in the surface region where the carrier density is high, as well as in the bulk of the medium where there is a low carrier density. The observed redshift in the PL emission with increasing growth temperature is attributed to hydrostatic relaxation of compressive strain, which is developed during the growth because of lattice parameter as well as the thermal mismatch between the substrate and grown material [41]. Usually, the strain developed by the native defects is hydrostatic in nature. In the present case, relaxation of compressive strain takes place with the formation of native defect comprising N vacancy. The redshift of $A_1$(LO) phonon mode (Figures 1(d)-1(f)) in the Raman spectra clearly indicates the relaxation of compressive strain with increasing growth temperature.

Unlike thin films it is very difficult to measure the carrier density directly, as our samples are nanoparticles in nature. One can calculate the carrier density using the BM shift [42],

$$\Delta_{BM} = \frac{\hbar^2}{2m^*}(3\pi^2 n_e)^{2/3} \qquad (3)$$

where $m^*$ is the reduced mass of the electron-hole pair, $\Delta_{BM}$ is the energy shift from the original band gap and $n_e$ is the carrier density in the system. For the calculation of carrier density using BM shift, the value of 0.75 eV is taken as intrinsic band gap of InN and $m^*$ is $0.052 m_o$ (electron effective mass, $m_e^* = 0.06 m_o$ and hole effective mass, $m_h^* = 0.44 m_o$) [43]. The shift is calculated from high emission point in the PL spectra, which show the BM shift as 0.45 eV. The carrier density, $n_e$ from the BM shift is calculated to be $2.53 \times 10^{19}$ cm$^{-3}$ using Equation (3). The calculated values of carrier density represent the bulk carrier density. However, one can expect the sheet carrier density to be more than an order of the bulk carrier density because of the SEA [39]. Presence of high amount of SEA, thus can originate plasmon peak in the THz frequency [34, 35].

On the other hand, PES technique is a well-established technique for studying the SEA. The UV photoemission spectroscopy (UPS) technique is employed to study the valence band spectrum (VBS) of InN nanostructures for the analysis of SEA because of its surface sensitivity. Figure 3(a) shows the VB spectra collected from the samples A, B and C. The position of valence band onset with respect to the Fermi level ($E_F$) is estimated from the intersection of the baseline and extrapolation line of the leading edge of the VB spectrum [44]. The methodology of determining the $E_F$ is explained in the SI. The positions of the valence band onset of the samples A, B and C are estimated to be 1.11, 2.90 and 3.09 eV below the $E_F$, respectively (Figure 3(a)). However, the VB spectra of InN are shadowed (especially for the sample C) by the O related emission (~6 eV), which may originate from the $Al_2O_3$ substrate as the nanostructures are grown non-uniformly on the substrate (Figures 1(a)−1(c)). It is worth mentioning that we have not observed any peaks corresponding to indium oxide in the Raman studies (Figures 1(d)−1(f)). So the role of O in the InN sub-lattice may be ruled out in the VB spectra. However, we found almost monotonous blue shift of VB spectra with increasing growth temperature. The shift cannot be related to O, as the substrate is same for all the samples. Thus, the blue shift in the VB spectra for samples with different growth temperatures is essentially the characteristic of InN (Figure 3(b)). In the present case, since the band gap of InN is in the range of 0.8-1.1 eV (Figures 1(g)-1(i)), the positions of valence band onset of the samples indicate that the $E_F$ is pinned into the conduction band. The pinning of the $E_F$ into the conduction band results in the downward surface band bending with the accumulation of surface electrons in these samples (SI, schematic Figure S4).

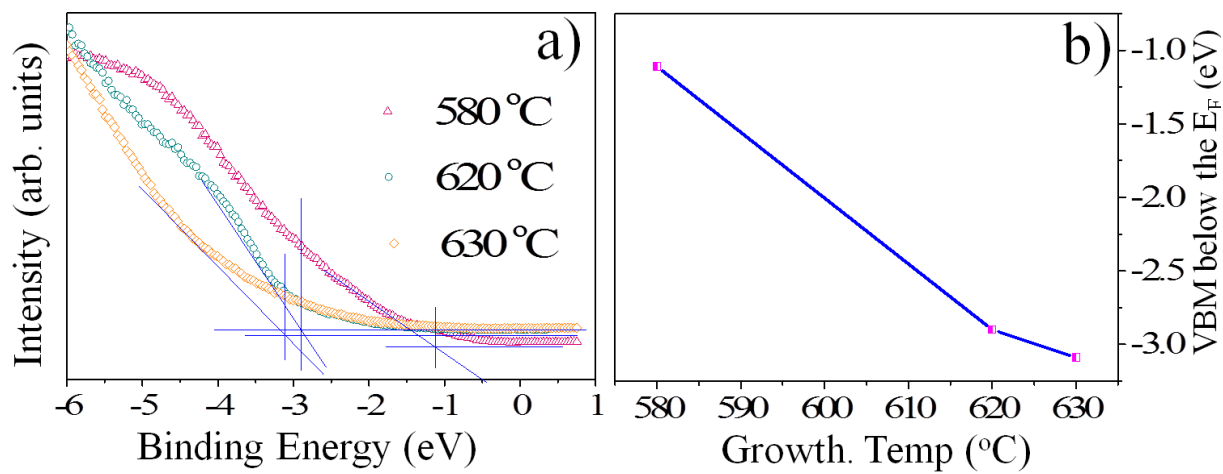

**Figure 3.** a) VB photoemission spectra of sample A (580 °C grown), B (620 °C grown) and C (630 °C grown). b) Pinning of Fermi level ($E_F$) into the conduction band shown increasing with increasing growth temperature. The line shown is a guide for eyes only.

**3.2 NSOM imaging and its correlation to LSPR and SPP**

It is well known that the propagating electromagnetic waves and the SPPs always have momentum ($k$) mismatch in their dispersion curves [1, 5]. To overcome the momentum mismatch, aperture NSOM probe is used to excite the SPPs. The light passing through the sub-wavelength aperture possesses higher spatial frequencies, which are required to excite SPPs [5]. These spatial frequencies, however, are not present in the propagating light. The range of spatial frequencies ($1/a$ to $1/\lambda$), available at the aperture, depends on the size of the aperture ($a$) and distance from it. Thus, the resolution of optical microscopy is increased by utilizing the evanescent field produced by a sub-wavelength aperture in the near-field. The resolution, however, depends on the size of aperture not with the wavelength of light [45]. In the present report, isolated nanoparticles with size up to 50 nm are imaged with an aperture size of 100 nm.

Topographic (Figures 4(a)–(c)) and corresponding NSOM images (Figures 4(d)–(f)) of nanostructures are shown for the sample A. The images, shown in figure 4(a) and 4(b), correspond to the two different areas of the sample A. However, image shown in figure 4(c) is the zoomed scan of the figure 4(b) (dashed area). The scale shown in NSOM images represents the counts per second of a single photon counter for a given integration time. The topography and NSOM images allow to directly correlate the optical intensity distribution with the structural image. The correlation between the topography and the NSOM images improved in zoomed scan because of effective scan rate is increased. The observed optical contrasts in the NSOM images are attributed to the intrinsic properties of individual nanoparticles. In case of scattered type NSOM with noble metal nanoparticle at the end of the tip, the image can be understood based on the quasi-static approximation. The amplitude of the scattering intensity depends on the effective polarizability of the tip, which has the contribution from the tip as well as the image dipole in the sample [46]. However, in the present case imaging of the nanostructures is performed using the aperture probe with light emitting from the aperture being considered as the point light source [47]. In the aperture case, there is always the same dipole defined by the aperture and the alternations in the contrast are due to changes in the dielectric properties of the sample or with the dipolar angle of a molecule interacting with the constant dipole [48].

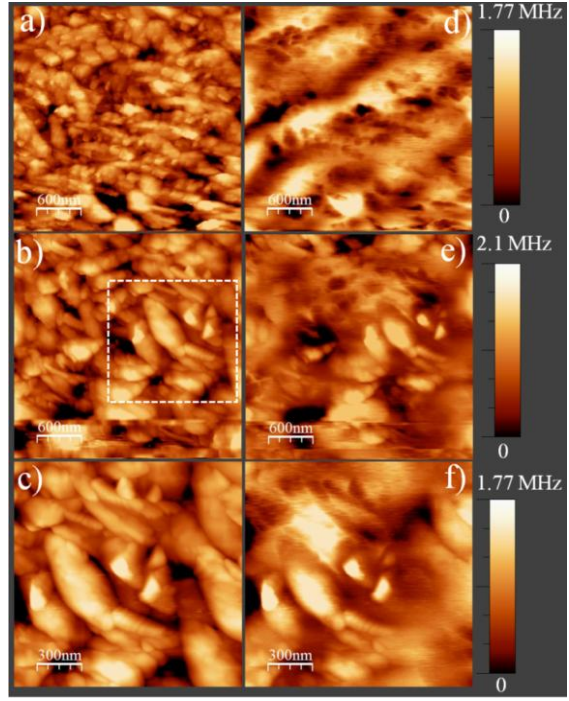

**Figure 4.** (a, b, c) Topography of sample A and their corresponding (d, e, f) NSOM images, respectively. Observed optical contrasts in the NSOM images attributed to variation in the carrier density of individual nanoparticles. Raw data is shown in the supporting information (SI, Fig. S5).

Thus, one can think that the sample polarizability is the sole reason for the observed optical contrast in the NSOM image and the near-field scattering from the individual nanoparticles depends on the local dielectric function of the sample ($\varepsilon_s$). In case of the doped semiconductor, the dielectric function of the sample can be written as $\varepsilon_s(\omega) = \varepsilon_{Opt} + \varepsilon_{Drude}$, where $\varepsilon_{opt}$ is the contribution from bound electrons and $\varepsilon_{Drude}$ is the Drude's contribution to the dielectric function [49]. The Drude part of the dielectric function is attributed the free carrier density in the system and the complex dielectric function of the sample can be further written as,

$$\varepsilon_s(\omega) = \varepsilon_{opt}\left(1 - \frac{\omega_p^2}{(\omega^2 + i\gamma\omega)}\right) \quad (4)$$

$$\omega_p^2 = \frac{ne^2}{\varepsilon_{opt}\varepsilon_o m_e^*} \quad (5)$$

where $\omega$ is the excitation frequency; $\omega_p$ is the plasma frequency which is related to free carrier density ($n$), and electronic charge ($e$); $\varepsilon_o$ is the static dielectric constant and $\gamma$ is the damping constant [49]. Since VBS shows that there is a negligible amount of SEA for the sample A (Figure 3(a)) and Raman spectroscopic study does not show any plasmon related peak (Figure 1(d)). Thus, sample A is considered as a lightly doped semiconductor as compared to those for the samples B and C. The free carrier density available in the system is attributed to the background carrier density instead of the SEA. For simplicity, if we neglect the damping parameter in Equation (4), dielectric function is solely dependent on $\omega_p$ and $\omega$. In case of lightly doped semiconductors, $\omega_p$ is significantly lower than the $\omega$ (2.32 eV in the present study). There can be free carrier absorption of light for $\omega > \omega_p$. Thus, the optical contrast in NSOM image (Figure 4) is attributed to free carrier absorption of the nanoparticles. The contrast in the NSOM images, however, depends on the carrier density of the individual nanoparticles. Nanoparticles, which are having high carrier density show significant absorption due to available free carriers [49]. Thus, with the variation of local carrier density there is a variation in absorption and hence in the scattered light intensities leading to the variation in the optical contrast of NSOM images, as it appears in sample A. To corroborate the assumption of variation in the carrier density, we collected the Raman spectra at different location of the sample A (SI, Figure S6). The background of each spectrum shows variation in slopes. Variable background of Raman spectra revealed the fact that there could be a variation in intrinsic carrier density of each nanoparticle [50]. The variation in the intrinsic carrier density is attributed to the local growth conditions, structural defects, chemical impurities and internal strain in the particles.

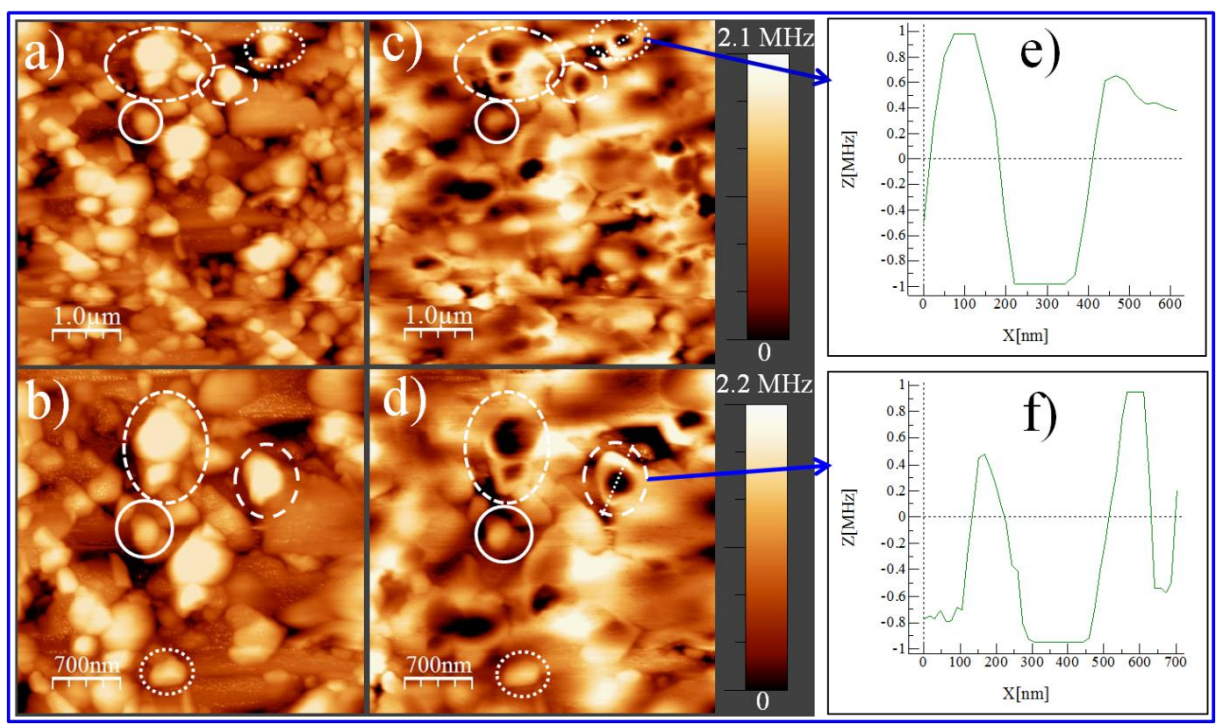

**Figure 5.** (a) Topography and corresponding (c) NSOM images of the sample B. NSOM images showing the strong extinction of the light at some of the nanoparticles and near-field enhancement around them. (b) The zoomed scan of (a) and (d) being the corresponding NSOM image showing similar kind of topography and optical images. The correlated particles in topography and NSOM images are shown with circles and ellipses. (e) and (f) showing the line profile of scattered intensity across the nanoparticles indicated by the blue arrows. Raw data is shown in the supporting information (SI, Fig. S7).

Topographic (Figures 5(a), 5(b)) and NSOM (Figures 5(c), 5(d)) images are shown for the sample B (grown at 620 °C). In contrast to the sample A, a significant absorption was observed in these nanoparticles. Line profiles of the scattered intensities are shown in figures 5(e) and 5(f). There is a significant enhancement of the electric field intensity surrounding each nanoparticle along with the strong absorption at the same time. In case of the sample B, the distinct optical contrast in the NSOM images may be because of strong absorption of the light. Enhancement of the electric field surrounding each nanoparticle is attributed to the resonance kind of behavior of nanoparticles with the availability of sufficient free carrier density originating from the SEA in the system. The plasmonic behaviour of these nanoparticles is also confirmed by the Raman spectroscopic analysis (Figure 1(e)). The SEA behaves like absolute free carriers similar to metals supporting plasmonic features.

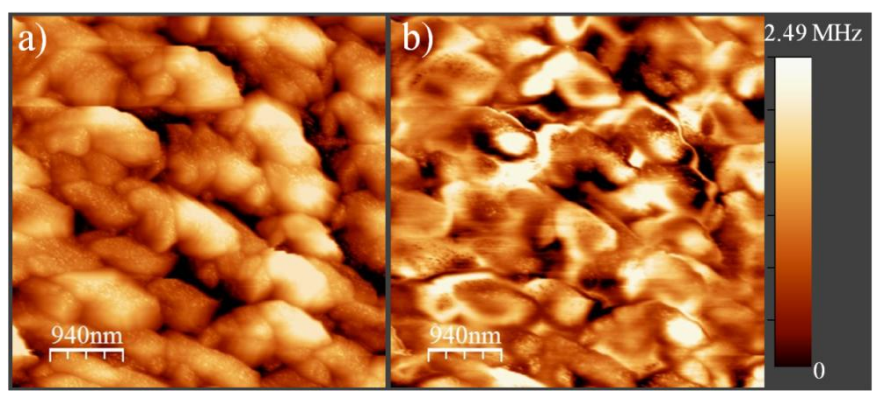

**Figure 6.** (a) Topography and corresponding (b) NSOM images of the sample C. Similar kind of near-field enhancement around each nanoparticle, as in case of sample B, being observed in the NSOM image except for a prominent absorption around nanoparticles. Raw data is shown in the supporting information (SI, Fig. S8).

From the VBS of sample C, the surface sheet carrier density is found to be increased as compared to those for the samples A and B (Figure 3(a)). In addition, plasmon peak in the Raman spectrum is enormously strengthened because of the increase of surface sheet carrier density in sample C (Figure 1(f)). As a result, the surface plasmon effects are expected to be more prominent in sample C. Figure 6 shows the topography (Figure 6(a)) and corresponding NSOM (Figure 6(b)) images of sample C collected at an arbitrary position. Similar kind of the optical contrast is also observed as in case of sample B (Figures 5(c) and 5(d)) with near-field enhancement of electric field leading to prominent absorption around each nanoparticle.

Figures 7(a) and 7(b) reveal the topography and corresponding NSOM images of the sample C, respectively, scanned at another arbitrary area. Three dimensional (3D) NSOM image corresponding to the figure 7(b) is shown in figure 7(d). However, there is a striking

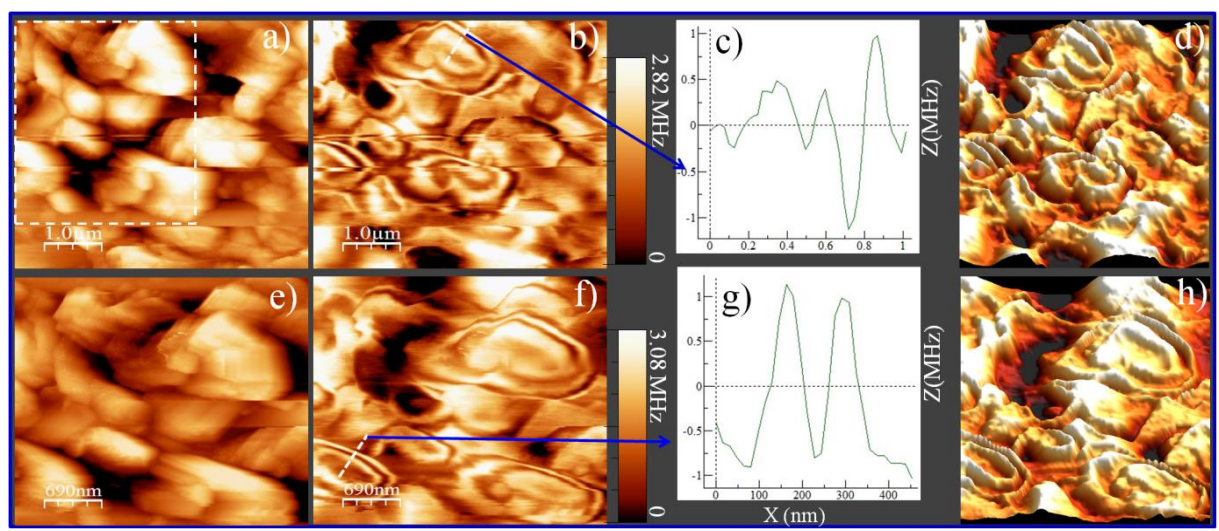

**Figure 7.** (a, e) Topography and corresponding (b, f) NSOM images, respectively for the sample C (630 °C). (e, f) zoom in scan of (a, b); zoomed area is indicated by dashed lines. Periodic fringes are clearly observed because of the interference of generated and reflected SPPs. (c, g) line profiles of periodic fringes. The collected region is indicated by the blue arrows. (d, h) 3D images corresponding to (b, f). Additional 3D images are also shown (SI, Figure S9) in the supporting information. Raw data is shown in the supporting information (SI, Fig. S10).

difference of the NSOM images of sample C as compared to the other samples. Periodic fringes are observed at each particle in the NSOM images instead of resonance kind of behaviour. The observed fringes are attributed to the excitation of the propagating SPPs (wavelength, $\lambda_p$). The evanescent field produced by the sub-wavelength aperture excites the SPPs and they propagate towards the edge of the surface where they get reflected back. Thus, the observed fringes may be the result of the interference of the excited and back reflected

SPPs [18]. Further, the topography of zoomed scan (dashed area in figure 7(a)) is recorded (Figure 7(e)). Similar kind of fringes is also observed in the corresponding NSOM image (Figure 7(f)) along with its 3D view (Figure 7(h)). Thus, it may be suggested that fringes observed in NSOM images are because of excitation of SPPs and not from any topography related artifacts. The line profile of the fringes corresponding to selected nanoparticles is shown in the figures 7(c) and 7(g). The periodicity of fringe ($\delta$), which is related to the wavelength of SPPs ($\delta = \lambda_p/2$), is calculated as 250 nm and 137 nm for the figure 7(c) and 7(g), respectively. Thus, the $\lambda_p$ is calculated as the 500 nm and 274 nm for the figures 7(c) and 7(g), respectively. The variation in the SPP wavelength for different nanoparticles is attributed to the variation in the carrier density of individual nanoparticles [51]. Origin of the SPPs must be attributed to the 2DEG of the SEA. This is further corroborated by the observed 2D plasmon peak in the Raman spectra (Figures 1(e) and 1(f)). Two-dimensional nature of electron gas and very high sheet carrier density are the sole reasons for the observation of the very small SPP wavelengths. Similar kind of shortening of SPP wavelength around forty times as compared to the excitation wavelength is reported in case of graphene [18, 19]. Moreover, in the present case, the possibility of excitation of SPPs with high excitation energy is attributed to the unique nature of band bending at InN surface. The surface downward band bending prevails, as the $E_F$ is pinned into conduction band. The $E_F$ is pinned to 2.9 eV above the valence band maximum near surface region (Figure 3(a)), which forbid the interband transitions with a given excitation of 2.32 eV (532 nm). Moreover, sheet carrier density possesses the metallic nature because of 2D nature of electron accumulation. Finally, the generation and stability of SPP are attributed to 2D plasmonic nature and unique surface band bending.

In order to confirm that the fringes observed are because of wave nature of SPPs, NSOM imaging is further performed in another area of sample C which shows the topography (Figure 8(a)) and corresponding NSOM (Figure 8(b)) images. However, from the topographic images, it is revealed that nanoparticles are within clustered regions. Interestingly, the NSOM images show interference fringes between particles.

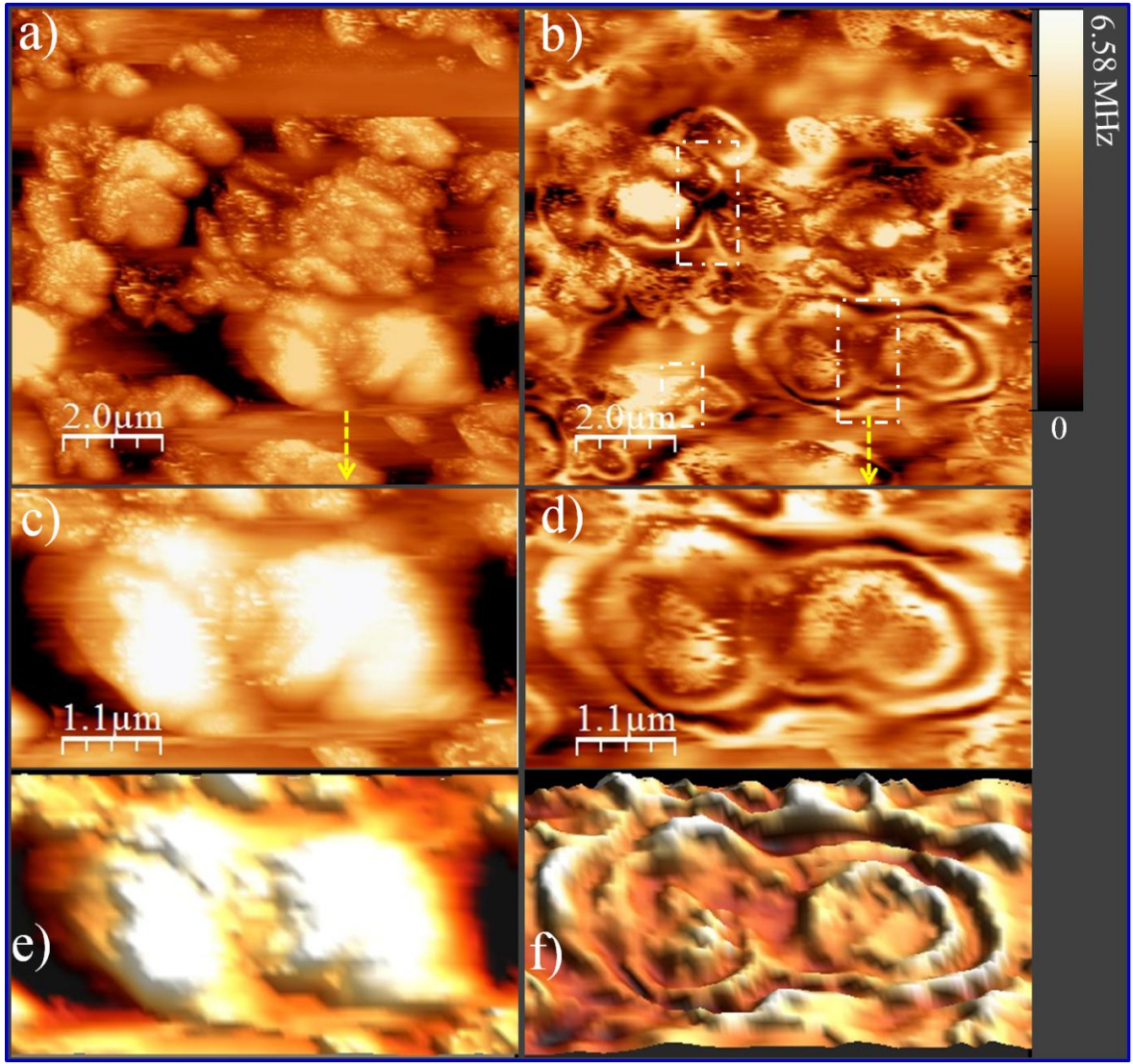

**Figure 8.** (a) Topography and corresponding (b) NSOM images of nanostructures of the sample C. Destructive interference of SPPs (in the dash-dotted area) generated by two adjacent InN nanoparticles is clearly observed. (c) Topography and corresponding (d) NSOM images show the zoomed image of the region revealing destructive interference (indicated by arrow marks). Corresponding 3D images are also shown in (e) Topography and (f) NSOM images. Raw data is shown in the supporting information (SI, Fig.S11).

These fringes are distributed in a way such that bright regions are seen at the edges and dark regions even at the top of the particle where there is the largest height in the topography. Thus, the contrast in the optical image cannot correlate to the topographic height effect of the nanostructures as it shows opposite character.Thus, the possible likely explanation is that the SPPs are generated by the cluster of nanoparticles which acts as a continuous medium. The interaction is clearly supported by the observed destructive interference between SPPs generated by the two adjacent nanoparticles in the cluster, as shown clearly in the dash-dotted areas (Figure 8(b)). In essence, the observation of the destructive interference further confirms that the recorded fringes are because of the SPPs generated by the accumulated surface electron of InN and not related to any other phenomena. The zoomed image of one of the regions revealing destructive interference (indicated by arrow mark) is shown in the topography (Figure 8(c)) and corresponding NSOM (Figure 8(d)) images. Corresponding 3D views are also shown for its topography (Figure 8(e)) and NSOM (Figure 8(f)) images.

In the present case, even though nanostructures possesses SEA, the SPP generation is not observed in every nanostructure as observed in the NSOM images of samples grown >620 $^{o}$C . In order to support the SPP propagation, nanostructure must have sufficient plane area on the surface normal to the excitation. In case of small nanostructures (size <$\lambda_p$), resonance enhancement around the nanoparticles is observed instead of generation of the SPP (Figure 5 and Figure 6). The observation of resonance enhancement is attributed to the generation of localized modes when the surface area less than $\lambda_p$. Similar kind of resonance enhancement is also observed in case of the tapered graphene nanostructures [19]. As a result, we observed the SPP generation in few of the NSOM images where the nanostructures have sufficient plane area on the surface. The large area FESEM image of sample C is shown in SI (Figure S12). The encircled areas show the nanostructures with large facets with area > (600×400) nm$^2$. Thus, the observed SPP is attributed to one of those large faceted nanostructures. This is further corroborated by the topographic image of AFM shown in figures 7(a) and 7(e).

**Conclusions**

As a summary, we explored the plasmonic properties of InN nanostructures by the real space imaging of the surface plasmons, for the first time, using the near-field scanning optical microscopy (NSOM). Plasmonic properties of InN are attributed to the two dimensional electron gas (2DEG) of surface electron accumulation (SEA). For this purpose, InN nanostructures were grown with varying sheet carrier density by varying the growth temperature. Variation in the surface sheet carrier density enables tailoring of the plasmonic properties of InN. The frequency of 2DEG corresponding to SEA is found to be in the THz region. Samples with the presence of SEA show resonance kind of behavior demonstrating near-field enhancement of light, as well as the propagating surface plasmon polaritons (SPPs) in NSOM images. The wavelength of SPPs, generated by InN nanoparticles, is estimated to be in the range of 274-500 nm. Shortening of SPP wavelength is attributed to 2D nature of the electron gas in SEA. Thus, InN can be a potential plasmonic material in THz plasmonic region, for which plasmonic properties can be tailored using the variation of carrier density in the SEA. Well organized InN nanostructures can even be utilized in the sub-wavelength photonic devices.

## ASSOCIATED CONTENT

### Supporting Information

GIXRD patterns are shown along with schematic band-diagrams of InN considering accumulation of surface electron. Raman spectra collected at different area of the sample A along with additional 3D NSOM images of the sample C and line profile where the SPP is observed. In addition to this, methodology of finding the surface electron accumulation using the different experimental techniques is also described.

## AUTHOR INFORMATION

### Corresponding Author

*E-mail: madupu@igcar.gov.in. Phone: +91-9445979580. Fax: +91-44-27480081.

*E-mail: dhara@igcar.gov.in. Phone: +91-9445603001. Fax: +91-44-27480081.

[#] Presently at Government Higher Secondary School, Pazhayannur-680587, Kerala, India

### Notes

The authors declare no competing financial interest.

### Acknowledgements

We would like to thank Mr. S. R. Polaki, SND, IGCAR for his help in FESEM studies. We thank Amirtha Pandian, Condensed Matter Physics Division, IGCAR, for their help with the HRTEM studies. We also thank Aaron Lewis, Nanonics Imaging Ltd., Israel for his useful discussion and suggestion in improving the content of the manuscript.

**References:**

[1] Maier S A 2007 *Plasmonics: Fundamentals and Applications* (Springer: New York)

[2] Atwater H A 2007 The promise of plasmonics *Sci. Am.* **296** 56–62

[3] Barnes W L, Dereux A and Ebbesen T W 2003 Surface plasmon subwavelength optics *Nature* **424** 824–30

[4] Homola J 2008 Surface Plasmon Resonance Sensors for Detection of Chemical and Biological Species *Chem. Rev.* **108** 462–93

[5] Novotny L, and Hecht B 2006 *Principles of Nano-Optics* (Cambridge University Press: New York)

[6] Gramotnev D K and Bozhevolnyi S I 2010 Plasmonics beyond the diffraction limit *Nat. Photonics* **4** 83–91

[7] Schuller J A, Barnard E S, Cai W, Jun Y C, White J S and Brongersma M L 2010 Plasmonics for extreme light concentration and manipulation *Nat. Mater.* **9** 193–204

[8] Boltasseva A and Atwater H A 2011 Low-Loss Plasmonic Metamaterials *Science* **331** 290–91

[9] West P R, S. Ishii S, Naik, G V, Emani, N K, Shalaev V M, and Boltasseva A 2010 Searching for better plasmonic materials *Laser Photonics Rev.* **4** 795–808

[10] Lindquist N C, Nagpal P, McPeak, K M, Norris D J and Oh S 2012 Engineering metallic nanostructures for plasmonics and nanophotonics *Rep. Prog. Phys.* **75** 036501.

[11] Luther J M, Jain P K, Ewers T, and Alivisatos A P 2011 Localized surface plasmon resonances arising from free carriers in doped quantum dots *Nat. Mater.* **10** 361–65

[12] Faucheaux J A, Stanton A L D, and Jain P K 2014 Plasmon Resonances of Semiconductor Nanocrystals: Physical Principles and New Opportunities *J. Phys. Chem. Lett.* **5** 976–85

[13] Mattox, T M, Ye X, Manthiram K, Schuck P J, Alivisatos A P and Urban J J 2015 Chemical Control of Plasmons in Metal Chalcogenide and Metal Oxide Nanostructures *Adv. Mater.* **27** 5830–37

[14] Lounis S D, Runnerstrom, E L, Llordés A and Milliron, D J 2014 Defect Chemistry and Plasmon Physics of Colloidal Metal Oxide Nanocrystals *J. Phys. Chem. Lett*. **5** 1564–74

[15] Ju L, Geng B, Horng J, Girit C, Martin M, Hao Z, Bechtel H A, Liang X, Zettl A, Ron Shen Y and Wang F 2011 Graphene plasmonics for tunable terahertz metamaterials *Nature Nanotechnology* **6** 630–34

[16] Koppens F H L, Chang D E and Abajo F J G 2011 Graphene Plasmonics: A Platform for Strong Light-Matter Interactions *Nano Lett*. **11** 3370–77

[17] Fei Z, Andreev G O, Bao W, Zhang L M, McLeod A S, Wang C, Stewart M K, Zhao Z, Dominguez G, Thiemens M, Fogler M M, Tauber M J, Castro-Neto A H, Lau C N, Keilmann F and Basov D N 2011 Infrared Nanoscopy of Dirac Plasmons at the Graphene-SiO2 Interface *Nano Lett*. **11** 4701–05

[18] Fei Z, Rodin A S, Andreev G O, Bao W, McLeod A S, Wagner M, Zhang L M, Zhao Z, Thiemens M, Dominguez G, Fogler M M, Castro Neto A H, Lau C N, Keilmann F and Basov D N 2012 Gate-tuning of graphene plasmons revealed by infrared nano-imaging *Nature* **487** 82–85

[19] Chen J, Badioli M, Alonso-Gonzalez P, Thongrattanasiri S, Huth F, Osmond J, Spasenovic M, Centeno A, Pesquera A, Godignon P, Elorza A Z, Camara N, Abajo F J G, Hillenbrand R and Koppens F H L 2012 Optical nano-imaging of gate-tunable graphene plasmons *Nature* **487** 77–81

[20] Allen Jr S J, Tsui D C and Logan R A 1977 Observation of the Two-Dimensional Plasmon in Silicon Inversion Layers *Phys. Rev. Lett*. **38** 980–83

[21] Olego D, Pinczuk A, Gossard A C and Wiegmann W 1982 Plasma dispersion in a layered electron gas: A determination in GaAs-(AlGa)As Heterostructures *Phys. Rev. B* **26** 7867–70

[22] Fasol G, Mestres N, Hughes H P, Fischer A and Ploog K 1986 Raman Scattering by Coupled-Layer Plasmons and In-Plane Two-Dimensional Single-Particle Excitations in Multi-Quantum-Well Structures *Phys. Rev. Lett*. **56** 2517–20

[23] Pietro P D, Ortolani M, Limaj O, Gaspare A D, Giliberti V, Giorgianni F, Brahlek M, Bansal N, Koirala N, Oh S, Calvani P and Lupi S 2013 Observation of Dirac plasmons in a topological insulator *Nature Nanotechnology* **8** 556–60

[24] King P D C, Veal T D, Jefferson P H, Hatfield S A, Piper L F J, McConville C F, Fuchs F, Furthmüller J, Bechstedt F, Lu H and Schaff W J 2008

Determination of the branch-point energy of InN: Chemical trends in common-cation and common-anion semiconductors *Phys. Rev. B* **77** 045316

[25] Mahboob I, Veal T D, McConville C F, Lu H and Schaff W J 2004 Intrinsic Electron Accumulation at Clean InN Surfaces *Phys. Rev. Lett*. **92** 036804

[26] Colakerol L, Veal T D, Jeong H, Plucinski L, DeMasi A, Learmonth T, Glans P, Wang S, Zhang Y, Piper L F J, Jefferson P H, Fedorov A, Chen T, Moustakas T D, McConville C F and Smith K E 2006 Quantized Electron Accumulation States in Indium Nitride Studied by Angle-Resolved Photoemission Spectroscopy *Phys. Rev. Lett.* **97** 237601

[27] Linhart W M, Chai J, Morris R J H, Dowsett M G, McConville C F, Durbin S M and Veal T D 2012 Giant Reduction of InN Surface Electron Accumulation: Compensation of Surface Donors by Mg Dopants *Phys. Rev. Lett*. **109** 247605

[28] Fu S P, Yu C J, Chen T T, Hsu G M, Chen M J, Chen L C, Chen K H and Chen Y F 2007 Anomalous Optical Properties of InN Nanobelts: Evidence of Surface Band Bending and Photoelastic Effects *Adv. Mater*. **19** 4524–29

[29] Bhuiyan A G, Hashimoto A and Yamamotoa A 2003 Indium nitride (InN): A review on growth, characterization, and properties *J. Appl. Phys*. **94** 2779–2808

[30] Polyakova V M and Schwierz F 2006 Low-field electron mobility in wurtzite InN *Appl. Phys. Lett*. **88** 032101

[31] Zhao S, Fathololoumi S, Bevan K H, Liu D P, Kibria M G, Li Q, Wang G T, Guo H and Mi Z 2012 Tuning the Surface Charge Properties of Epitaxial InN Nanowires *Nano Lett*. **12** 2877–82

[32] Shubina T V, Ivanov S V, Jmerik V N, Solnyshkov D D, Vekshin V A, Kopev P S, Vasson A, Leymarie J, Kavokin A, Amano H, Shimono K, Kasic A and Monemar B 2004 Mie Resonances, Infrared Emission, and the Band Gap of InN *Phys. Rev. Lett*. **92** 117407

[33] Shubina T V 2010 Plasmonic effects and optical properties of InN composites with In nanoparticles *Phys. Status Solidi A* **207** 1054–61

[34] Shubina T V, Andrianov A V, Zakharin A O, Jmerik V N, Soshnikov I P, Komissarova T A, Usikova A A, Kopev P S, Ivanov S V, Shalygin V A, Sofronov A N, Firsov D A, Vorobev L E, Gippius N A, Leymarie J, Wang X

and Yoshikawa A 2010 Terahertz electroluminescence of surface plasmons from nanostructured InN layers *Appl. Phys. Lett*. **96** 183106

[35] Shubina T V, Gippius N A, Shalygin V A, Andrianov A V and Ivanov S V 2011 Terahertz radiation due to random grating coupled surface plasmon polaritons *Phys. Rev. B* **83** 165312

[36] Madapu K K, Polaki S R and Dhara S 2016 Excitation dependent Raman studies of self-seeded grown InN nanoparticles with different carrier concentration *Phys. Chem. Chem. Phys*. **18** 18584–89

[37] Davydov V Yu, Emtsev V V, Goncharuk I N, Smirnov A N, Petrikov V D, Mamutin V V, Vekshin V A and Ivanov S V 1997 Experimental and theoretical studies of phonons in hexagonal InN *Appl. Phys. Lett*. **75** 3297–99

[38] Miroshnichenko A E, Flach S and Kivshar Y S 2010 Fano resonances in nanoscale structures *Rev. Mod. Phys.* **82** 2257–98

[39] Veal T D, Piper L F J, Schaff W J and McConville C F 2006 Inversion and accumulation layers at InN surfaces *J. Cryst. Growth* **288** 268–72

[40] Chang Y, Mi Z and Li F 2010 Photoluminescence Properties of a Nearly Intrinsic Single InN Nanowire *Adv. Funct. Mater* **20** 4146–51

[41] Kisielowski C, Kruger J, Ruvimov S, Suski T, Ager III J W, Jones E, Liliental-Weber Z, Rubin M, Weber E R, Bremser M D and Davis R F 1996 Strain-related phenomena in GaN thin films *Phys. Rev. B* **54** 17745

[42] Walsh A, Da Silva J L F and Wei S 2008 Origins of band-gap renormalization in degenerately doped semiconductors *Phys. Rev. B* **78** 075211

[43] de Carvalho L C, Schleife A and Bechstedt F 2011 Influence of exchange and correlation on structural and electronic properties of AlN, GaN, and InN polytypes *Phys. Rev. B* **84** 195105

[44] Chambers S A, Droubay T, Kaspar T C and Gutowski M 2004 Experimental determination of valence band maxima for $SrTiO_3$, $TiO_2$, and SrO and the associated valence band offsets with Si(001) *J. Vac. Sci. Technol. B* **22(4)** 2205–14

[45] Dürig U, Pohl D W and Rohner F 1986 Near-field optical-scanning microscopy *J. Appl. Phys*. **59**, 3318–27

[46] Keilmann F, and Hillenbrand R 2004 Near-field microscopy by elastic light scattering from a tip *Phil. Trans. R. Soc. Lond. A* **362** 787–805.

[47] Ren X, Liu A, Zou C, Wang L, Cai Y, Sun F, Guo, G and Guo G 2011 Interference of surface plasmon polaritons from a "point" source *Appl. Phys. Lett.* **98** 201113

[48] Betzig E and Chichester R J 1993 Single Molecules Observed by Near-Field Scanning Optical Microscopy *Science* **262** 1422–25

[49] Fox M 2001 *Optical Properties of Solids* (Oxford University Press: New York)

[50] Inushima T, Higashiwaki M and Matsui T 2003 Optical properties of Si-doped InN grown on sapphire (0001) *Phys. Rev. B* **68** 235204

[51] Fei Z, Goldflam M D, Wu J -S, Dai S, Wagner M, McLeod A. S, Liu M K, Post K W, Zhu S, Janssen G C A M, Fogler M M and Basov D N 2015 Edge and Surface Plasmons in Graphene Nanoribbons *Nano Lett.* **15(12)** 8271–76

Supporting Information

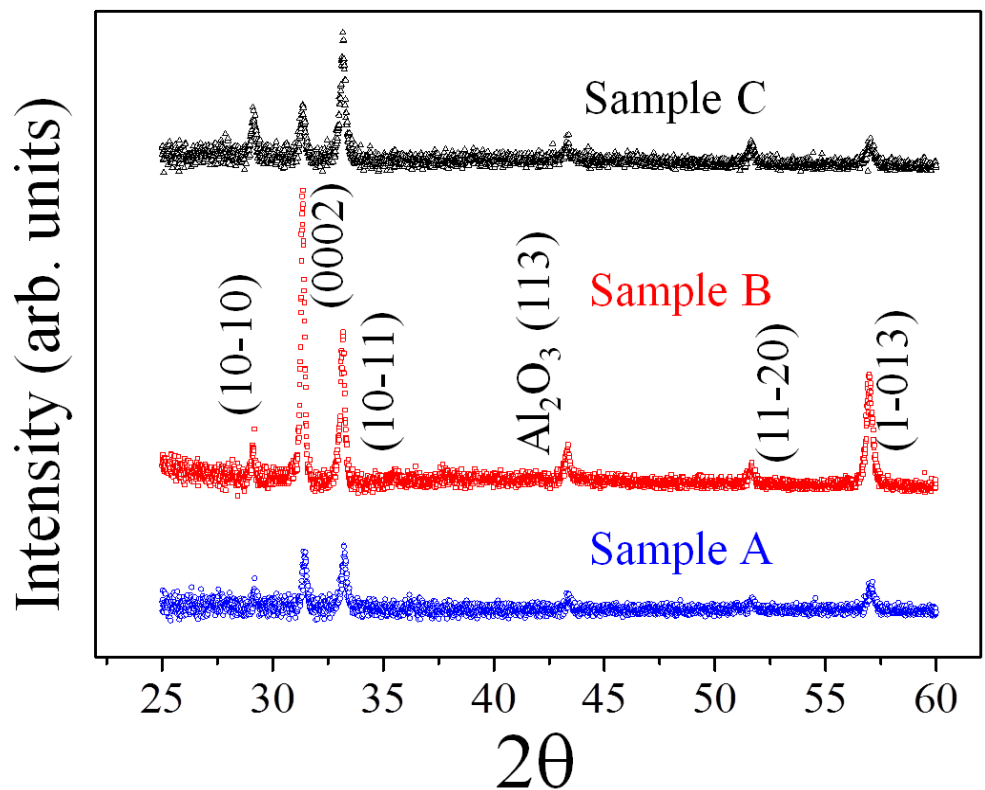

**Figure S1.** GIXRD patterns showing hexagonal phase (JCPDS card # 00-050-1239) of InN nanostructures grown at three different temperatures.

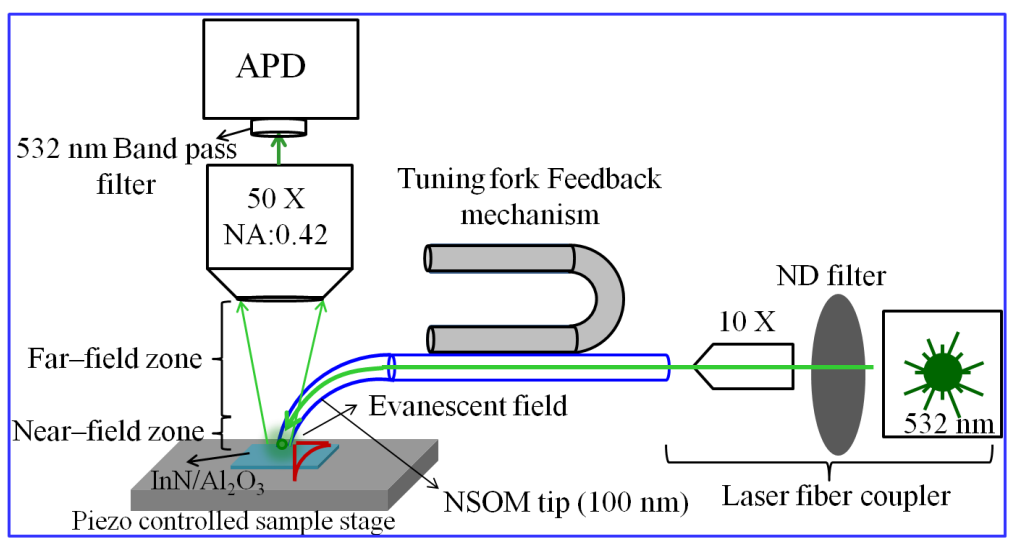

**Figure S2**. Schematic of NSOM technique used for the study of near-field optical properties of InN nanostructures.

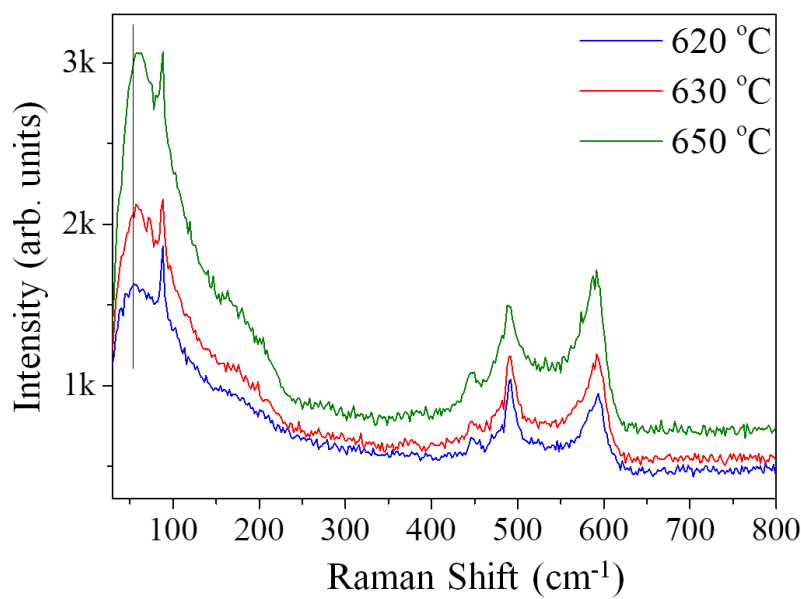

**Figure S3.** Variation of 2D plasmon peak freqency with increasing growth temperature.

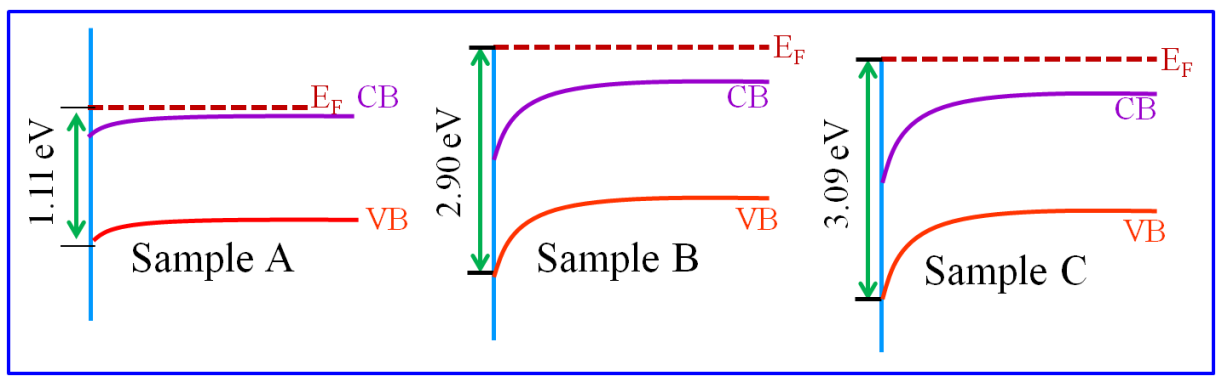

**Figure S4.** Schematic band diagrams for samples A, B and C illustrating the downward band bending of samples grown at high temperature. The band bending values are based on the estimates from UV photoemission spectroscopy (UPS) studies in Figure 3a of the manuscript and discussion there in.

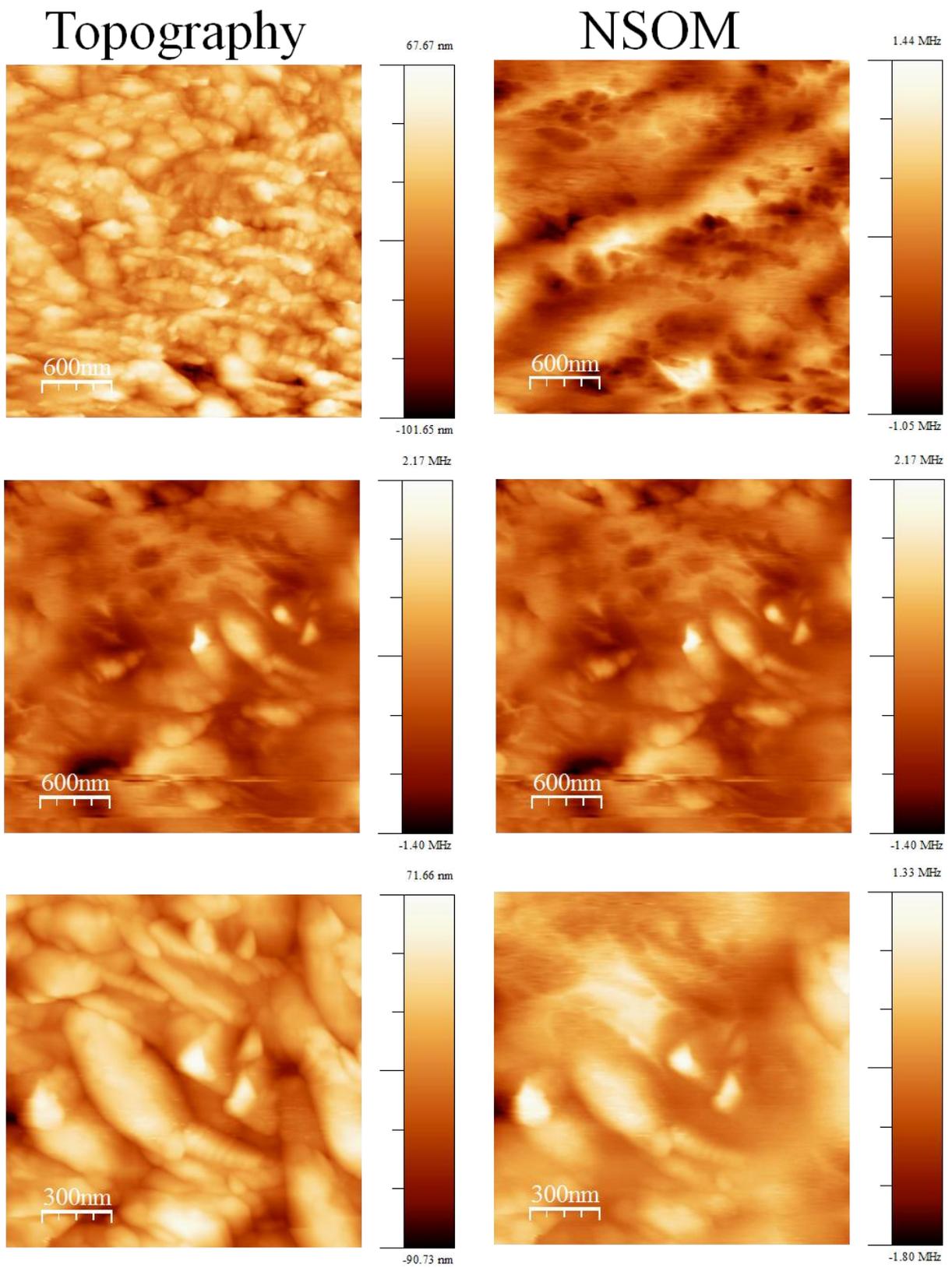

**Figure S5.** Raw data corresponding to Fig. 4

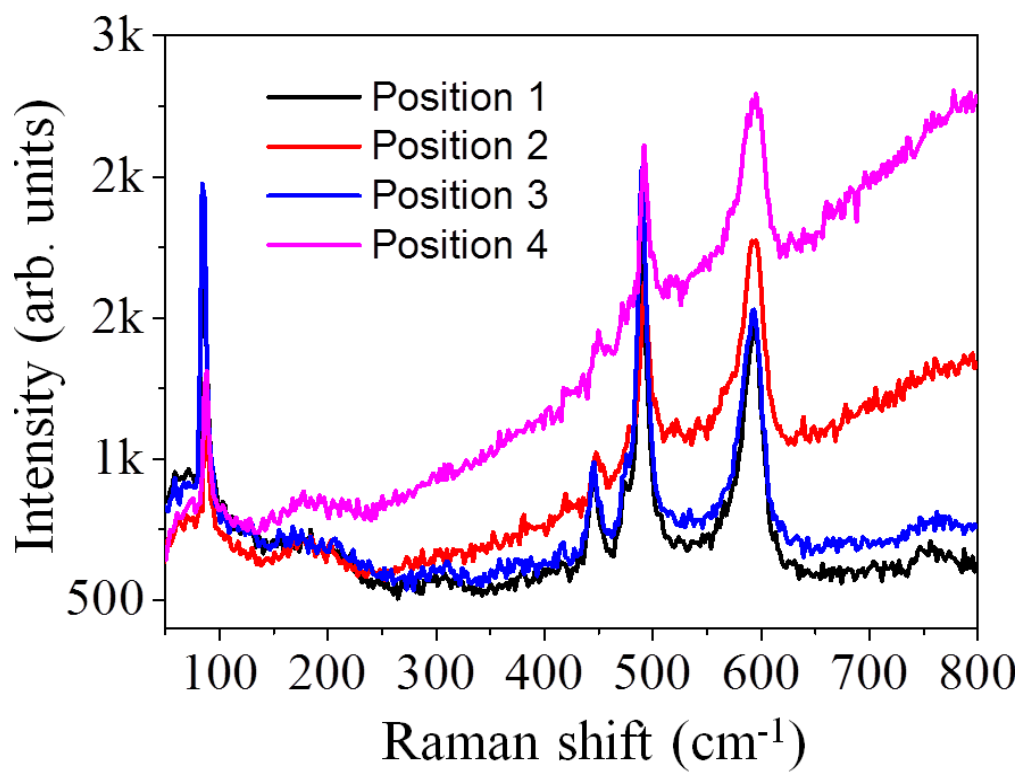

**Figure S6**. Raman spectra collected at arbitrary positions of the sample A (580 °C), showing the variable slope in the Raman background. Raman background qualitatively indicates about the amount of carrier density in the system. As the Raman spectrum collected using high numerical aperture (0.95) lens, the laser spot size (360 nm) covers barely one or two nanoparticles. In other words, there is a variable carrier density in the individual nanoparticles which is revealed from the observed Raman spectra collected at different spots in the micrograph.

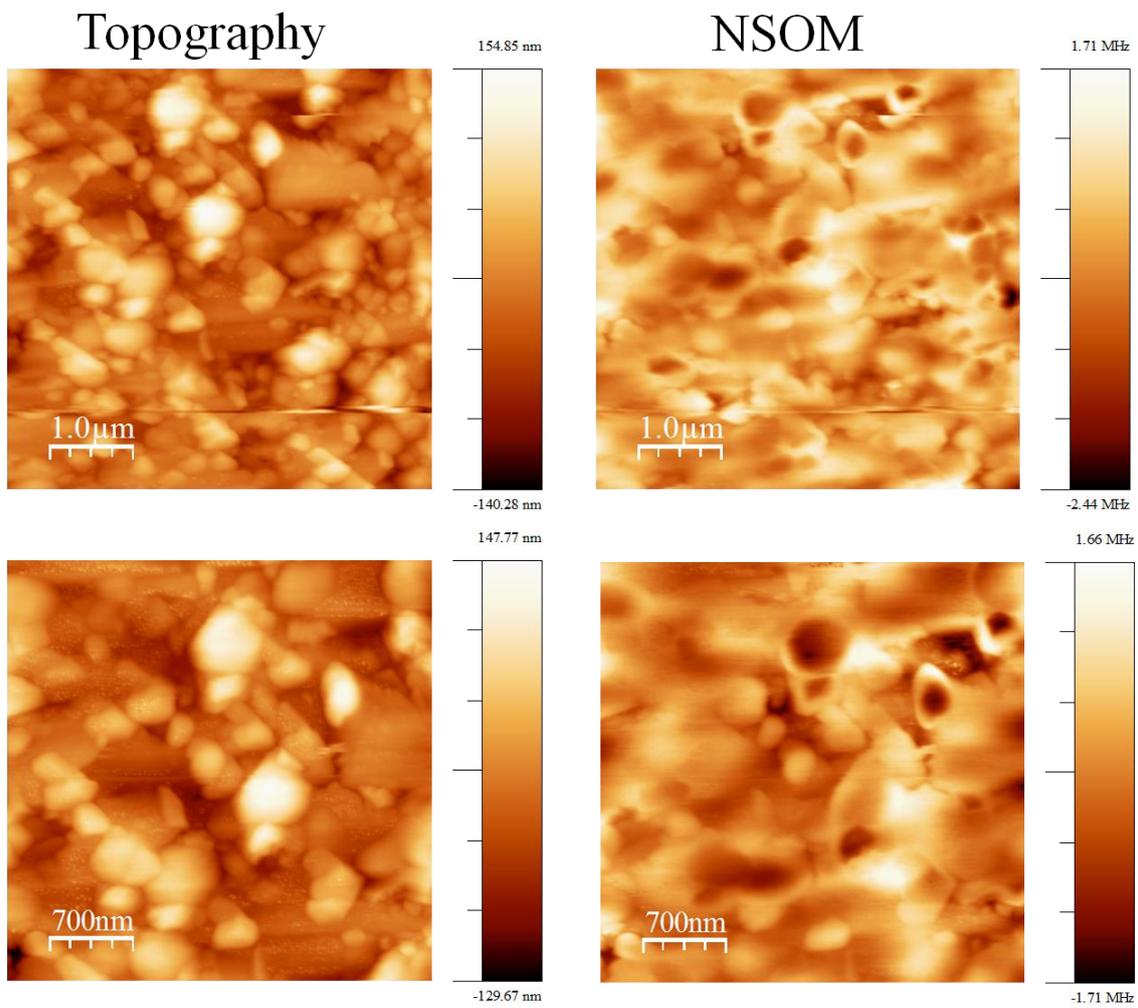

**Figure S7.** Raw data corresponding to Fig. 5

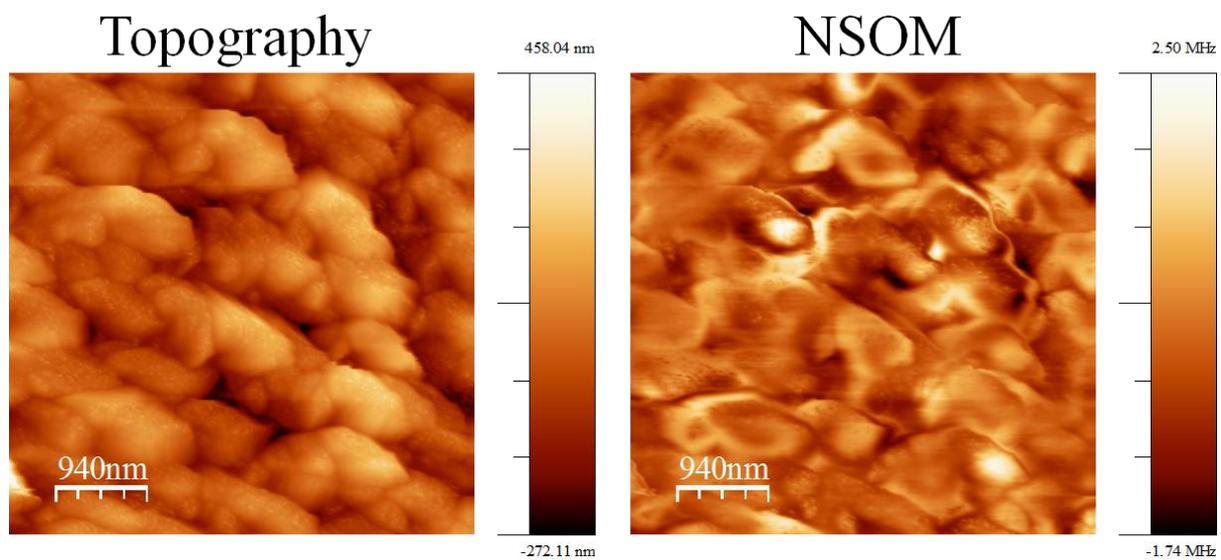

**Figure S8.** Raw data corresponding to Fig. 6

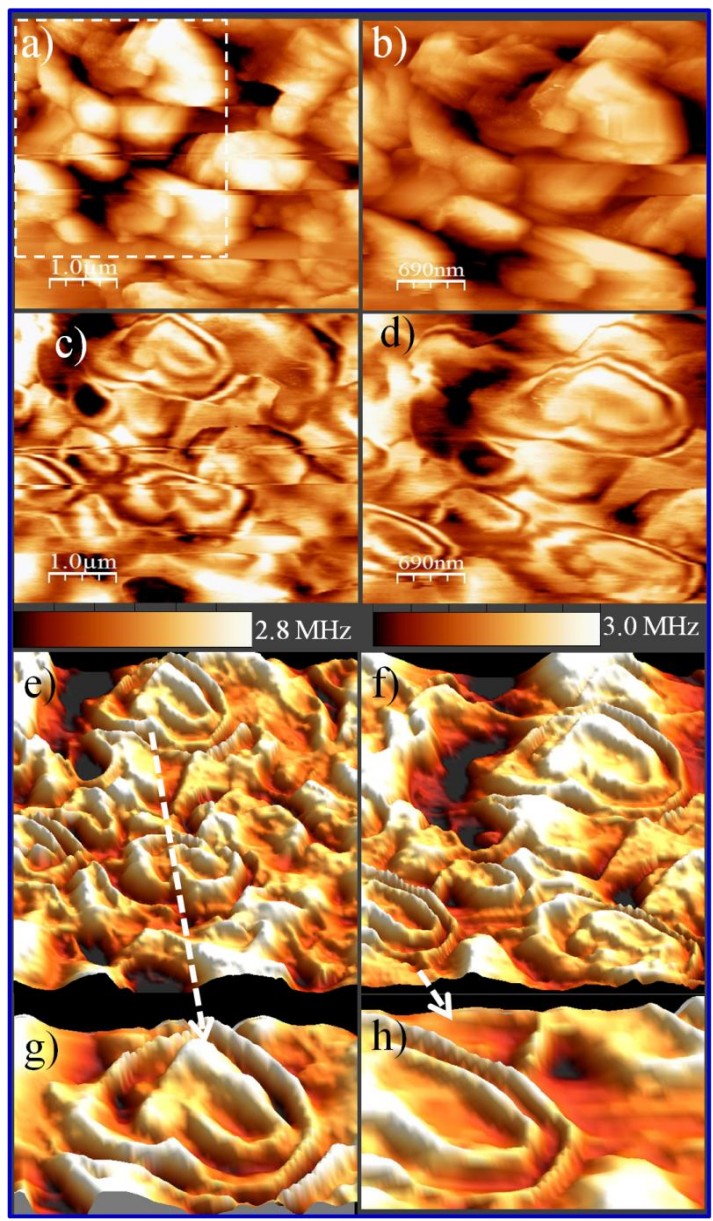

**Figure S9.** Topography (a, b) and, corresponding NSOM (c, d) and their respective 3D images (e, f), respectively for sample C (630 °C). Figure (b) is zoomed scan of (a) corresponding to the dashed area. Figures (g, h) are the zoomed 3D NSOM images of the single particles, indicated by arrow marks.

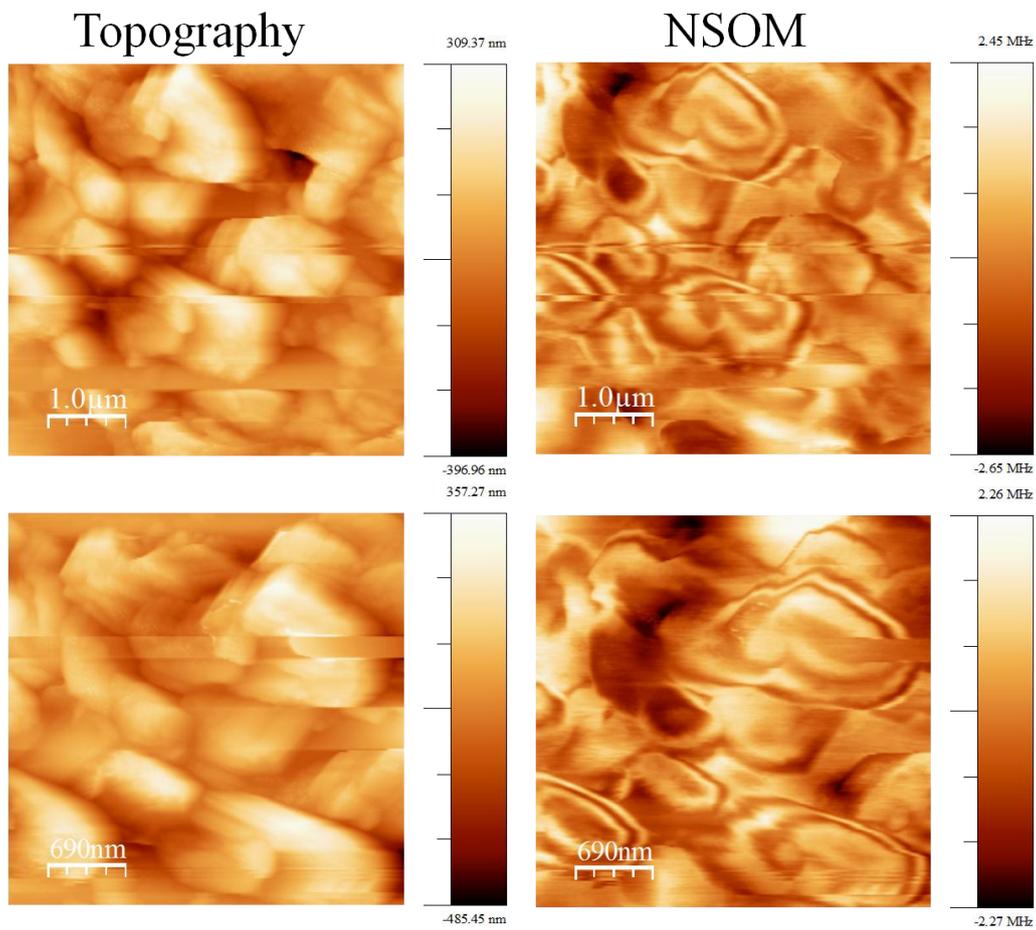

**Figure S10.** Raw data corresponding to Fig. 7

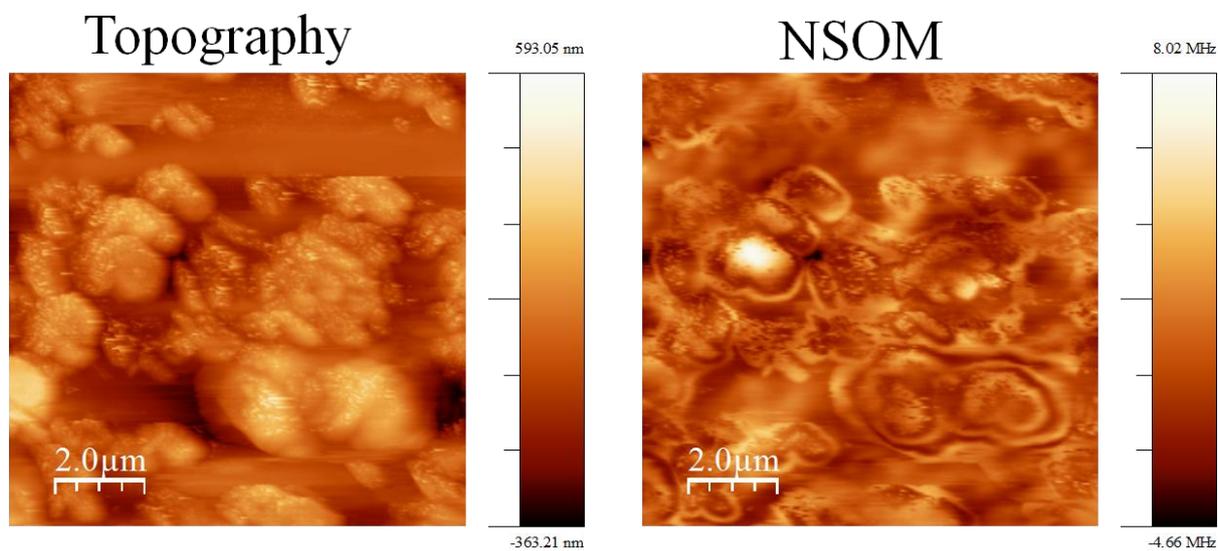

**Figure S11.** Raw data corresponding to Fig. 8

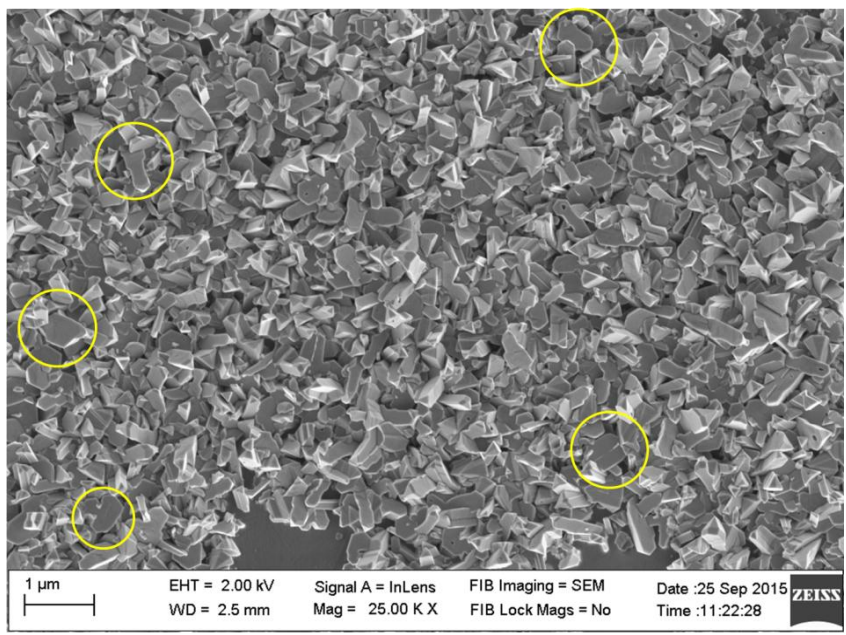

**Figure S12**. Large area FESEM micrograph of InN nanostructures grown at 630 °C. Encircled areas show large faceted nanostructures.

**Methodology for the determination of surface electron accumulation**

Existence of surface electron accumulation (SEA) in InN is proved by several experimental techniques.[1–3] However, high resolution electron loss spectroscopy (HREELS) and photoemission spectroscopy (XPS or UPS) are proven to be efficient techniques for studying the surface electronic properties. In the former technique of HREELS, one can study electronic properties of the complete space charge region by varying incident electron energy. The evidence of electron accumulation appears from the red shift of plasmon related peaks in the electron energy loss spectra with increasing incident electron energy.[2] In the later case of XPS or UPS study, valence band spectra are collected with respect to the Fermi level ($E_F$) and the position of the valence band (VB) maxima is determined by extrapolating the leading edge of VB spectrum. SEA is inferred by the position of $E_F$. If the $E_F$ is located in the conduction band, there can be downward surface band bending with the accumulation of surface electrons.[2,3] Amount of surface sheet carrier density is inferred from the amount of Fermi level pinned into the conduction band. In addition to these, both Raman and PL spectroscopic studies also provide considerable information regarding the SEA in a simple and quick manner.[4,5]

*References*:

1. H. Lu, W. J. Schaff, L. F. Eastman, C. E. Stutz, *Appl. Phys. Lett.* **2003**, *82*, 1736.
2. I. Mahboob, T. D. Veal, C. F. McConville, H. Lu, W. J. Schaff, *Phys. Rev. Lett*. **2004**, *92*, 036804.
3. W. M. Linhart, J. Chai, R. J. H. Morris, M. G. Dowsett, C. F. McConville, S. M. Durbin, T. D. Veal, *Phys. Rev. Lett*. **2012**, *109*, 247605.
4. Y. Cho, M. Ramsteiner, O. Brandt, *Appl. Phys. Lett*. **2013**, *102*, 072101.
5. Y. Chang, Z. Mi, F. Li, *Adv. Funct. Mater*. **2010**, *20*, 4146.